\documentclass{article}
\usepackage{preamble}

\title{ Quantifying Collection Lag \\ in European Modern and Contemporary Art Museums
\vspace{-0.1cm}
}

\author{
Mar Canet Solà*‡\textsuperscript{1,}\textsuperscript{2} 
Antonina Korepanova*‡\textsuperscript{1,}\textsuperscript{2} \\
Ksenia Mukhina\textsuperscript{1,}\textsuperscript{3} 
Maximilian Schich\textsuperscript{1,}\textsuperscript{2} \\
\small \textsuperscript{1}ERA Chair for Cultural Data Analytics, Tallinn University  \hspace{1mm} \\
\small \textsuperscript{2}Baltic Film, Media and Arts School, Tallinn University  \hspace{1mm} \\
\small \textsuperscript{3}School of Digital Technologies, Tallinn University  \hspace{1mm} \\
\small ‡ Corresponding authors: mar.canet@tlu.ee, antonina.korepanova@tlu.ee \\
\small * equal contribution as first authors 
}

\date{\small May 14, 2023 \vspace{-0.4cm}} %
\begin{document}
\maketitle

\begin{abstract}
   Museum collection strategies are governed by a variety of factors, including topical focus, acquisition funds, availability of works in the art market, donations and specific coincidental opportunities. Yet, it remains unclear if more fundamental collection patterns emerge, exist, and are shared between museums, which could for example allow an established artist to estimate when a contemporary art museum would acquire their works. Here we collect and analyze data from 12 European contemporary art museums, taking into account artwork creation dates, collection acquisition dates, and the associated artist age at both points in time. From this simple quantitative construct we are able to reveal a striking gradient of museum profiles at the aggregate level. This lag can function to constitute a macroeconomic index of "mean museum collection lag", ranging from 3 years in the most dynamic cases (Kiasma) to 33 years in the most established institutions (Reina Sofia). Meanwhile, on the granular level, plotting artist age over collection year, and using artist-age vs artwork-collection matrices, a detailed picture becomes evident, where individual museums are characterized by shared patterns and a rich heterogeneity of ideographic details. Regularities include continuous acquisitions, systematic acquisition of older materials over time, and brief bursts, where whole oeuvres of individual artists join specific collections. Hence, we are able to shed light on the detailed collection history of museums, transcending the anecdotal nature of art historical storytelling via the provision of a quantitative context. Our approach of cultural data analysis combines expertise in art, art history, computational social science, and computer science. Our joint perspective builds a bridge between and serves an audience of museum professionals, art market actors, collectors, and individual artists alike.
\end{abstract}

\begin{figure}[t]
         \centering
         \includegraphics[width=\linewidth]{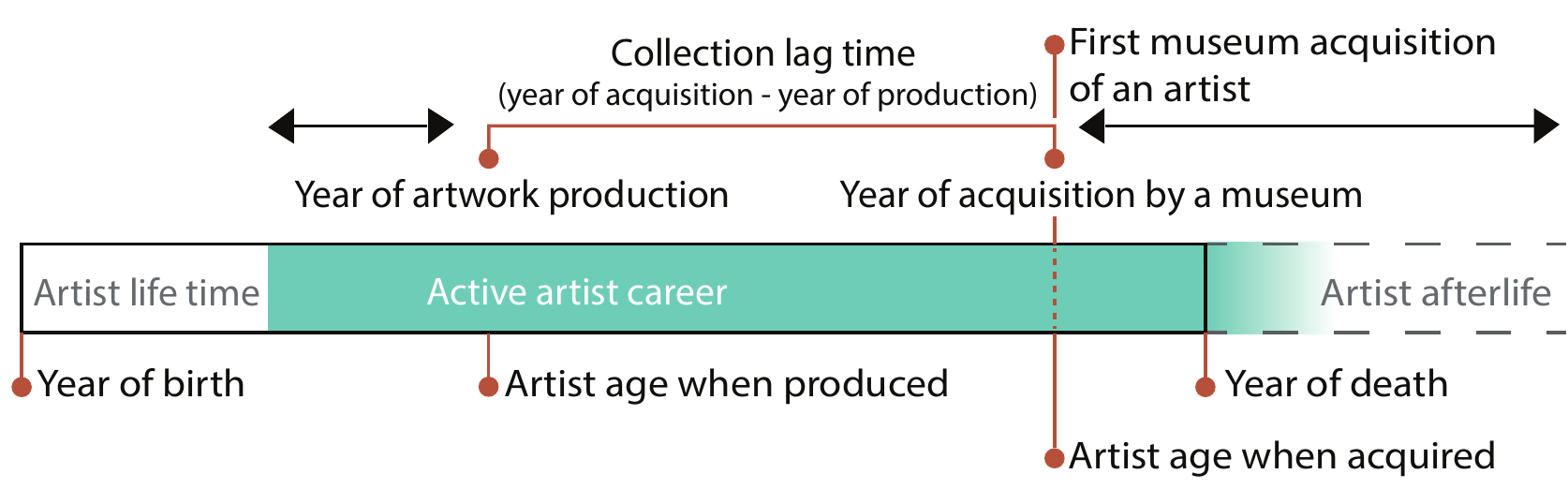}
         \caption{Schema of analyzed data points (red).}
  \label{fig_museum_data}
\end{figure}

\section{Introduction}

Museum collections enrich art appreciation and understanding in a broad public audience, while facilitating scholarship via the accumulation of relevant information. This study uses principles of cultural data analytics to explore such museum data, characterizing the art collection process in twelve contemporary art museums, using data visualization as a key method to provide insight into acquisition strategies that vary across museums. Salient aspects are the collection lag and artist career stage at time of collection. Relying on quantitative visualizations, our contribution transcends more traditional art research that takes a more qualitative perspective, focusing on individual artists, groups, or a single museum. An unprecedented availability of data makes our contribution possible. Our approach of combining methods of quantification and visualization with large sets of data from public institutions, resonates with other applications of Quantitative Art History and Cultural Analytics \parencite{reyes2020cultural,manovich2020cultural,Schich2017Getty, schich2016figuring}.

Alan Bowness - art historian and director of the Tate Gallery in the 1980s - observed in his lecture “The Conditions of Success” at the University of London \parencite{bowness_conditions_1990} that exceptional artists are getting recognized after approximately 25 years of active work, with museums playing an essential role in the latest steps of what the author describes as peer recognition.
Moreover it is recognized that art collections have a transformative role in shaping the meaning and value of objects \parencite{macdonald2006collecting}. This starts during the acquisition process of contemporary artworks since multiple stakeholders exhibit a vested interest in establishing the artwork's identity, while evaluating it for acquisition \parencite{ryan2016considerations}, based on their view of the object's cultural and aesthetic value. This process invariably leads to detailed discourse and evaluation from diverse perspectives \parencite{ryan2016considerations}. As objects are integrated into a collection, they acquire an added layer of significance by virtue of their membership in the collection \parencite{macdonald2006collecting, miller2022museum}, adding a historical and contextual perspective to an artist's work. This means the collection as a whole is literally more than the sum of its parts, essentially putting art research into the realm of complexity science \parencite{anderson1972more, IbrusSchichTamm2022}. In addition, collections are often formed with the intention of long-term or even perpetual preservation, thus creating a final stage in an object's biography, while simultaneously attempting to imbue it with a more enduring and meaningful existence. As such, objects in collections enter a new phase in their biographies \parencite{kopytoff1986cultural} marked by different cultural storage practices, cataloging, and display. This aspect highlights the intricate relationship between art collections and the objects they contain.

Consequently, the museum acquisition of artworks is a significant milestone in the career of many artists. It stimulates economic exchange in the art market and increases the value of the artworks created by the artist. We note that this study does not differentiate between the methods of acquisition, whether through purchase or donation. Second, acquiring artworks at different times in artists’ lives plays a crucial role; museums play either a supportive role in art-making (by acquiring artworks at the beginning of artistic careers) or a conservatory role (by acquiring artworks after the artists’ death or at later points of artistic careers). 

For this study, we define museum acquisition as the process of incorporating artworks or artefacts into a museum's permanent collection, as documented via the registration of object in the museum inventory. Such registration can be consequence of various processes, such as purchasing, gifting, bequesting, or donating by different actors like artists, art collectors, galleries, institutions or public actors such as the state. Feeding into the typical core museum mission of  collecting, preserving, and exhibiting cultural heritage, acquiring artworks or artefacts requires careful consideration and evaluation by museum professionals, including curators and conservators, who assess the objects' authenticity, provenance, condition, and significance. The underlying acquisition policies, conventions, and procedures are essential for museums as they govern the growth and diversity of their collections \parencite{topaz2019diversity}.  

In this research, we introduce the concept of "mean collection lag" in contemporary art museums, referring to the median time elapsed between artwork creation and acquisition for a given contemporary art collection. The lag intertwines with and reflects acquisition strategy, with implications for artist careers and object preservation. Too large a lag may for example challenge object preservation, risking premature destruction of cultural heritage. Tracked over time, the collection lag further can help to unveil different and changing museum acquisition strategies, unveiling implicit emerging institutional structures and specificities, further acknowledging that museums have never been neutral \parencite{burgess2021state}.

\begin{figure*}[t]
 	\noindent
 	\includegraphics[width=\textwidth]{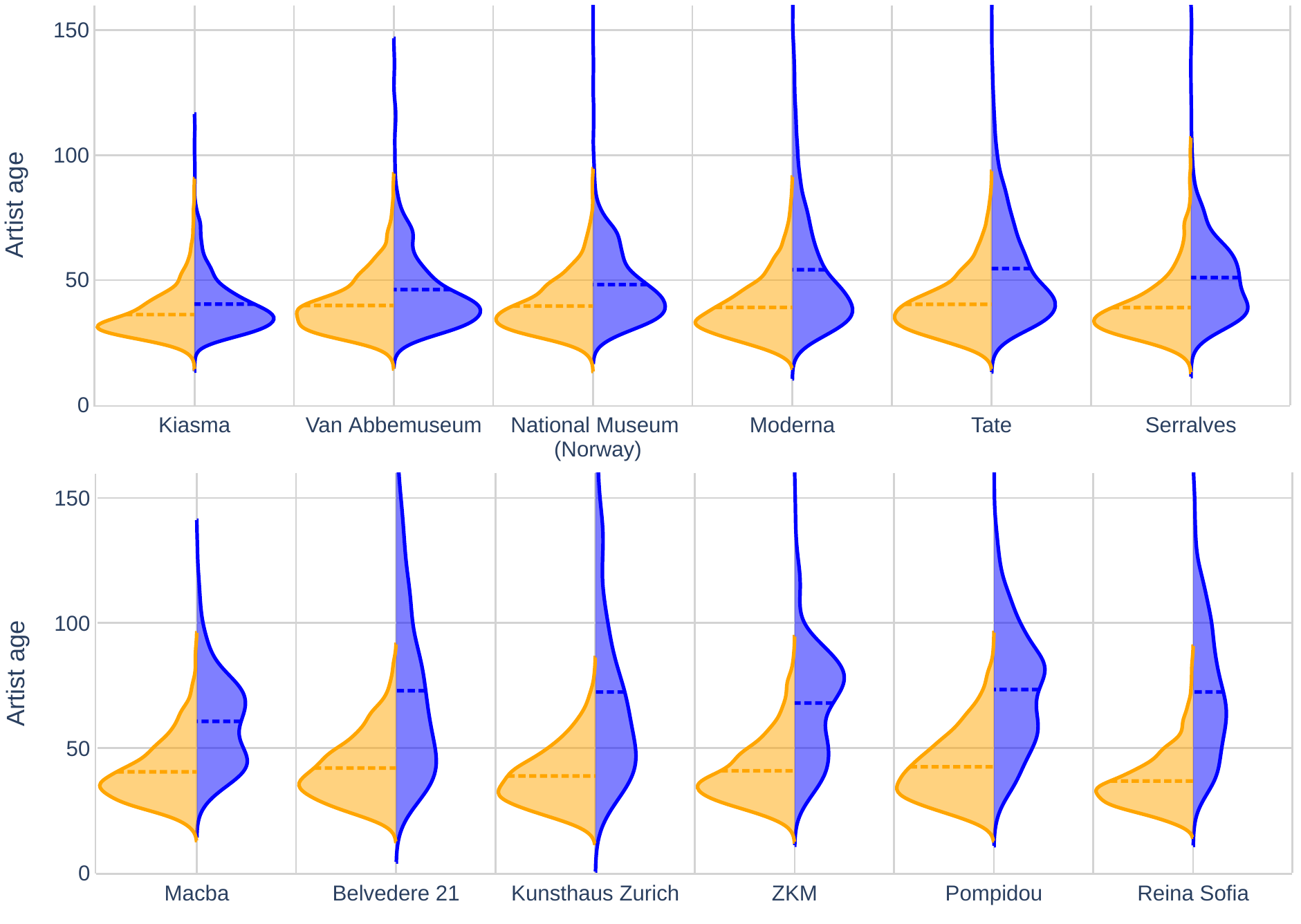}
 	\caption{ Split violins indicating density of artworks created (yellow) and acquired (blue) at artist age in 12 European museums. The difference of horizontal lines in each violin defined the median collection lag, increasing in museums left to right.
 	}
  \label{fig_violin}
 \end{figure*}

\section{Research context and questions}

In today's digital age, cultural data analysis has become a powerful tool for researchers, utilizing big data, computational, and visualization techniques to explore both current and past cultures \parencite{manovich2020cultural, IbrusSchichTamm2022}. Massive cultural data sets of visual content, such as digitized artworks, photos, movies, and interactive media \parencite{hristova2016images, poulopoulos2022digital} are analyzed to find patterns, trends, and insights beyond the scope of conventional approaches, opening new methods and forms of knowledge to the humanities \parencite{hall2013toward}, while harnessing deep rooted and shared foundations of art research, multidisciplinary science, and computation \parencite{schich2019cultural}. In line with such theories of cultural analysis, our study bridges computational methods with a topic traditionally researched by qualitative museum scholars, feeding into the development of a systematic multidisciplinary science approach to make sense of art and culture that uses quantification and visualization to clarify a diversity of meaningful perspectives.

Museum studies or museology critically examines the histories, theories, cultures, and politics of museums \parencite{latham2019whither}, where the act of collecting represents a particular intervention in the life of objects affecting their meaning and value \parencite{elsner1996new}. The study of collecting is a core topic within museum studies \parencite{kuo2015bibliometric}, where a wide variety of methods are used, including exhibition analysis, narratology, citation analysis, and sociology \parencite{sitzia2023museology}, gallery analyses, ethnographic fieldwork at museums, and interviews with museum visitors, professionals, and stakeholders \parencite{tucker2014museum}. Our study exemplifies that exploratory visual analysis \parencite{keim2006challenges, battle2019characterizing} functions as another valid method in this spectrum.

Museum inventory analysis is a promising direction to employ computational methods. Some public museums (MOMA, MET, Tate) prefer open their data to the public, thus stimulating interest in quantitative analyses of their collections \parencite{manovich2015data}. Additionally, many researchers clean and publish museum inventory-related datasets to sustain various directions of digital humanities and computer science research \parencite{knoblock2013publishing, szekely2013connecting, de2013amsterdam}. Greenwald \parencite{greenwald2022changing}, in related work, for example utilized a dataset on 18th and 19th-century French paintings, revealing that portraiture was the dominant genre and that there was an increase in female models over time. A later study \parencite{greenwald2019modernization} revealed the connection between landscapes and rural genre painting in France with the introduction of affordable train ticket prices from Paris. Quantitative analysis of the British Museum demonstrated that periods of active purchases for enriching the collection alternated with periods of donations \parencite{macdonald2022counting}. Another study, in a top-tier multidisciplinary science venue, \parencite{fraiberger2018quantifying} analyzed the exhibition history of half a million artists and found that early access to prestigious central institutions offered life-long access to high-prestige venues and reduced the dropout rate, while starting at the network periphery resulted in the opposite effect on artistic careers.

In our study, we surmised that looking deeply into the age of artists and artworks at the moment of acquisition might reveal emerging patterns in institutional acquisition strategies, resulting in the following questions: 

\begin{enumerate}
\item{How does age disparity between artists at the time of artwork production and museum acquisition shed light on the museums acquisition practices?}
\item{How does the acquisition pattern of living artists compared to deceased artists vary across museums, and what does this reveal about the level of support for living artists in these institutions?}
\item{How do the patterns of artists' ages at the moment of acquisition change over time, perhaps reflecting systematic trends in museum acquisitions?}
\item{How does the comparison between artist age at artwork acquisition versus artwork creation shed light on the structural differences across museums?}

\end{enumerate}

\section{Methodology}

\subsection{Museum selection and data acquisition}

When identifying museums to include in the study, we carefully considered several parameters. Firstly, we focused on top-tier institutions, as they have more resources and curatorial impact. Secondly, we selected museums that specialize in collecting modern and contemporary art, as they are pivotal agents in validating current artistic careers. We contacted a larger list of museums, eventually including all museums that agreed to share their data for our purpose. Thus, our sampling strategy falls into the category of purposeful sampling. In order to increase the representativeness of our sample, we attempted to include the largest and most influential museums from Europe, thereby employing a stratified purposeful sampling technique. In this paper, we focus on Europe, while future work will include museums worldwide.

For data collection, we employed several strategies. Initially, we used open datasets that were available on GitHub (Tate). Furthermore, some of the museums had a public Application Programming Interface, i.e. a set of protocols and tools that allowed us to gather data algorithmically (Kiasma, The Centre Pompidou). Next, we requested datasets directly from museums (National Museum - Norway, MACBA, Belvedere 21, Moderna - Sweden, ZKM, Kunsthaus Zürich, Serralves). Finally, we also scapred data from websites with Python scripts (Moderna - Sweden, Museo Reina Sofia). 

Overall, the participating museums were established in different centuries, which shapes their collections. Younger museums have mostly contemporary art in their datasets (Kiasma), while other museums collected modern and contemporary art (Serralves, ZKM, MACBA), while yet other museums also acquired artworks from previous times (Kunsthaus Zurich, Tate, Van Abbemuseum, Moderna - Sweden, Belvedere 21, The Centre Pompidou, National Museum - Norway, Reina Sofia).

\subsection{Data description and integrity}

Due to varying data standards and formats in each museum, data preparation and cleaning consisted of several steps. Some museums provided two tables: one with artwork metadata and another with artists' biographical data, which we joined in step one. All datasets were cleaned using the Python data-processing libraries Pandas and NumPy. To eradicate errors in calculations, several people worked with the same dataset and compared the results. We preserved artist names and artwork titles in original format. Numeric columns (acquisition date, creation of artworks, birth, and death of artists) were all transformed into a four-digit numbers indicating year. While extracting relevant years, in cases with two dates present in one cell for acquisition, the final date was defined according to artwork genre: for prints, photos, lithographs (and all other artworks that have several copies) we used the earlier year, otherwise the later year was used since the original production year not the reprints or reconstructions are relevant for our study.

The prepared data eventually contained six columns used in analysis: Artist year of birth, Artist year of death, Year of artwork production, Year of artwork acquisition, Artist name, and Artwork Label. Two fields were calculated based on existing data: Artist age and Artwork age at the moment of acquisition. A schematic description of the prepared data is presented in Figure \ref{fig_museum_data}.

The datasets originally included metadata exclusively concerning artworks held within the collections of the museums, with data presented in CSV, JSON, and XLS formats. However, due to the vast time frame of over 500 years and the predominantly manual nature of the digitization process, the original datasets contained some missing and inconsistent data. For example, the absence of information regarding certain artists made it impossible to obtain their birth and death years. Additionally, some typos in the data resulted in the year of acquisition being erroneously recorded as earlier than the year of creation. Some of the datasets have not been updated in recent years, such as Tate, which was last updated in October 2014. As a result, the visualizations produced from the Tate dataset may be outdated compared to those from other museums. Certain museums, such as ZKM, specialize in new media art, resulting in a collection of younger artworks, while museums containing traditional art forms, such as paintings and sculptures, may possess a larger number of older pieces like Kunsthaus Zürich founded in 1787 which is the oldest institution in our study. From this alone we may surmise that the dynamic in the acquisition of younger and living artists may be partially linked to the medium with which they work. Digital media and technology-related artwork has only recently emerged, implying most artists working with it are still alive. Finally, some datasets are a subset of an entire collection, such as those from The Centre Pompidou, Kunsthaus Zürich, and Moderna - Sweden, either partially digitized or only partially available to the public.

The datasets utilized in this research are subject to a Creative Commons license, which permits researchers to employ the data. The authors assert that the participating museums have not been misrepresented. In some cases we communicated electronically with museum representatives to get additional data in case of identified gaps or issues. A few datasets were acquired on the condition that they wouldn't be made available to the general public. Furthermore, some museums allowed data to be extracted directly from their websites because these sites had comprehensive information concerning their collections.

\begin{figure*}
 	\noindent
 	\includegraphics[width=\textwidth]{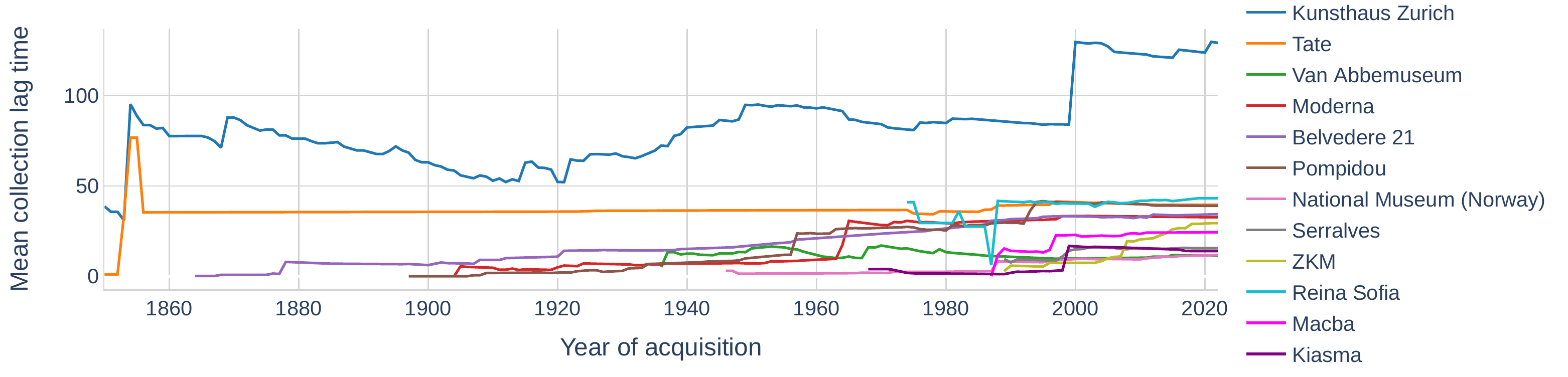}
 	\caption{Cumulative evolution of collection lag between artwork creation and acquisition in 12 European museums.}
  \label{fig_colectionLag}
\end{figure*}

To ensure methodological consistency, the study employed a filtering process to standardize the collections data. Specifically, the dataset was refined to exclusively incorporate works with a creation date later than 1860 in Figures \ref{fig_violin}, \ref{fig_colectionLag}, \ref{fig_ageAbsoluteTime}. This cutoff at 1860, of course somewhat arbitrarily, marks the onset of the Modern art period. Consequently, our study encompasses the era of classical modern and contemporary art. Additionally, the requirement for the creation age of artworks to exceed 18 years was implemented to address the issue of museums grappling with the registration of collective works. This filtering approach mitigates instances where some artworks were registered reflecting individual artist age, while others were registered reflecting the age of an artist collective as merely one or two years, combining artists that were of course not toddlers.

\section{Results}

\subsection{Gap between production and acquisition}
The violin plots in Figure \ref{fig_violin} visually depict the association between the age at which artists create their works and the age at which museums acquire them. The left side of the plot depicts the ages at which the artists created their works, whereas the right side of each plot displays the distribution of ages at which museums typically acquire artworks. The plots show that some museums acquire artworks considerably later than their creation date, as indicated by a higher mean on the right side of each plot. This is termed as "mean museum collection lag," which may reflect art creation stimulation and preservation strategies, with shorter gaps indicating faster acquisition of artworks after creation.

The gap between creation and acquisition is influenced by two factors: the age of the artist at the time of production and acquisition. Mean age at creation varies across museums, ranging from 34 (Kiasma) to 40 years (The Centre Pompidou, Belvedere 21), and is skewed towards younger ages, gradually dropping in numbers around 80 years. The mean age of acquisition varies more and spreads from 37 years (Kiasma) to 73 years (The Centre Pompidou). Some distributions are skewed towards older ages (ZKM, The Centre Pompidou) while others are closer to younger ages (National Museum - Norway, Moderna - Sweden, Tate, Belvedere 21). Some museums have an even distribution (Kiasma, MACBA, Museo Reina Sofia). Some museums show two bumps, indicating two waves of more intense acquisitions, one during the more typical for a mid-career artist and a second wave of acquisition close toward the end of the artist's life or shortly after death (MACBA, The Centre Pompidou, ZKM).

A variety of violin shapes emerge from the aforementioned differences; some plots (National Museum - Norway, Kiasma, Moderna - Sweden) are more symmetrical than others. The plots appear more asymmetrical the wider the difference between the median values on the left and right. The mean age difference ranges from 3 to 33 years (Kiasma and Museo Reina Sofia, respectively).
 
Looking at another aspect, Figure \ref{fig_colectionLag} indicates that the age of artist first acquisitions has increased over the years in all European museums under study, with the exception of the Tate, which in the 19th century acquired very old art and has decreased age of artist first acquisition then being flat in the last century. This finding sheds light on the changing dynamics of museum collecting practices and their implications for artist careers and the changing role of museums over time.

\begin{figure*}
 	\includegraphics[width=\textwidth]{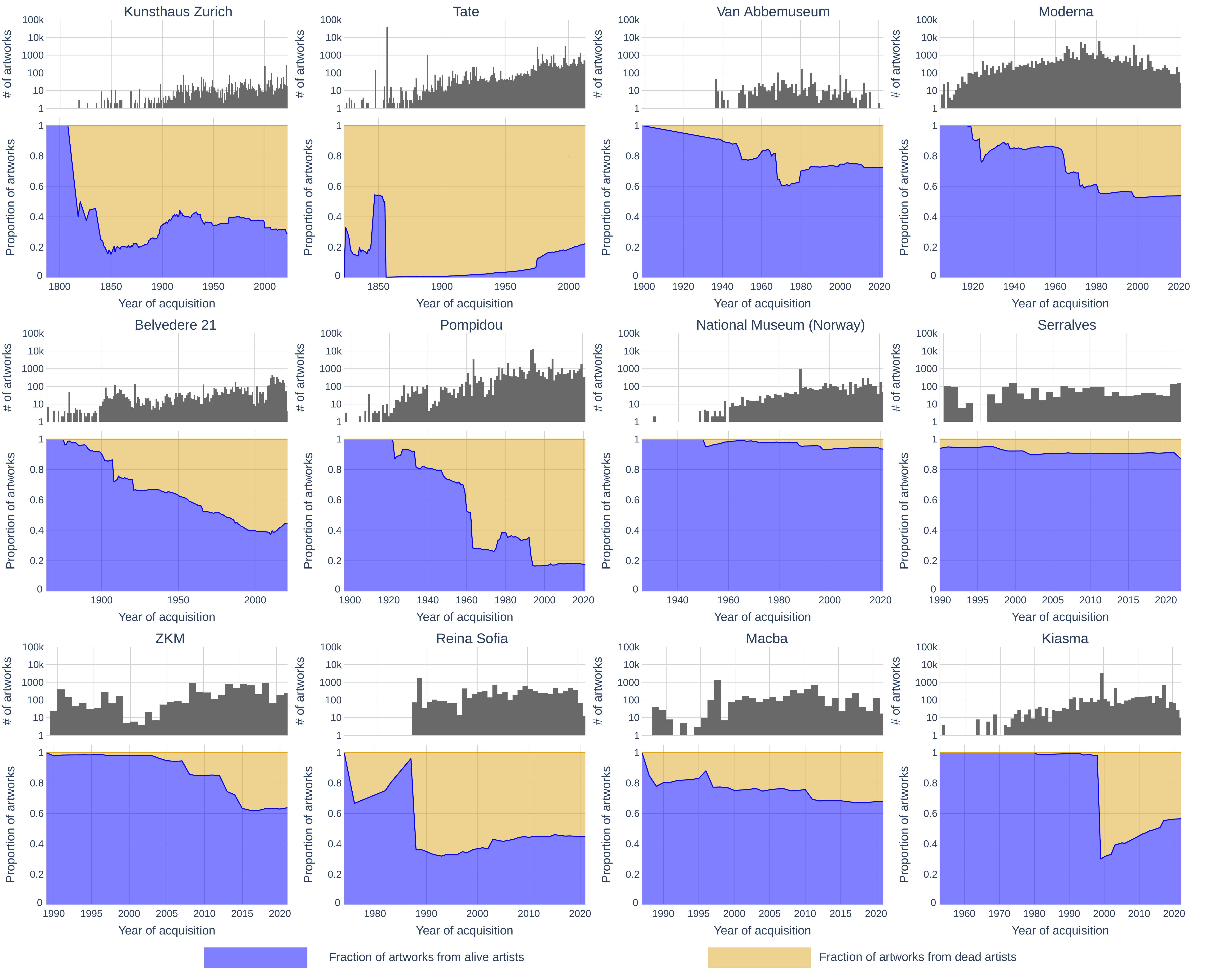}
 	\caption{Log acquisition number and cumulative alive/dead (blue/yellow) artist fraction in 12 European museums over time.   
 	}
  \label{fig_evoAliveDead}
 \end{figure*}
 
\subsection{The evolution of artist acquisition}
The evolution of the percentage of artworks acquired from living and deceased artists is a valuable metric that sheds light on the museum's values and priorities. The variation in living artist support across museums can be revealed through the analysis of this data. Figure \ref{fig_evoAliveDead} provides a comprehensive overview of the trend in alive versus dead artist acquisitions across several museums. The data shows that some museums have a more pronounced focus on acquiring works from deceased artists, while others have a more balanced approach.

Figure \ref{fig_evoAliveDead} contains 12 subplots arranged in a 4x3 grid, with each subplot representing a different museum ordered from oldest to youngest. Each subplot has an area chart displaying the trend of the percentage of artworks in the museum collection acquired of alive artists (blue) versus dead artists (blue) over time, and a bar chart displaying the number of artworks acquired per year. The main subplots (area charts) show the cumulative percentage of alive vs. dead acquisitions in a museum's collection over time. The horizontal axis measures time in years, and the vertical axis shows the percentage of artworks. Using cumulative percentages helps to identify important periods of growth or changes in the collection. Each subplot in the figure has a gray-colored bar chart at the top indicating the total number of acquisitions over time, with a logarithmically scaled y-axis to avoid skewing towards higher values. The x-axis of the bar chart corresponds to the x-axis of the area chart and shows the years of the museum's artwork collection.

Figure \ref{fig_evoAliveDead} reveals several patterns. Tate stands out as the only museum with a yellow-dominated graph, indicating that most artworks were acquired after the artist's death. However, this is largely due to a major Turner bequest in 1856 (and our choice of cumulating the measure). From 1856 to 1970, the Tate's collection remained stable at 0.99-0.94\% dead artists, but after the 1970s, the acquisition of artworks from living artists resulted in a gradual increase of the line to 0.77\% in 2013. Museo Reina Sofia and Kiasma also showed decreasing trends in the percentage of dead artists acquired (0.63\% to 0.55\% and 0.7\% to 0.43\%, respectively). Both museums had sudden surges in acquisitions of dead artists in 1988 and 1999, respectively, following years with only 0.03\% and 0.01\% dead artist acquisitions.

The opposite pattern might be noticed, with the percentage of dead artists' acquisitions gradually growing throughout the collection period, which may indicate that museums stick to their emerging program. While some museums demonstrate a considerable change of proportion (Moderna - Sweden, Belvedere 21, The Centre Pompidou, ZKM, Van Abbemuseum), other's dead/alive ratio appears to remain stable (National Museum - Norway, Serralves). 

\subsection{Trends in museum acquisitions}

\begin{figure}
 	\noindent
  \centering
 \includegraphics[width=\linewidth,height=0.99\textheight,keepaspectratio]{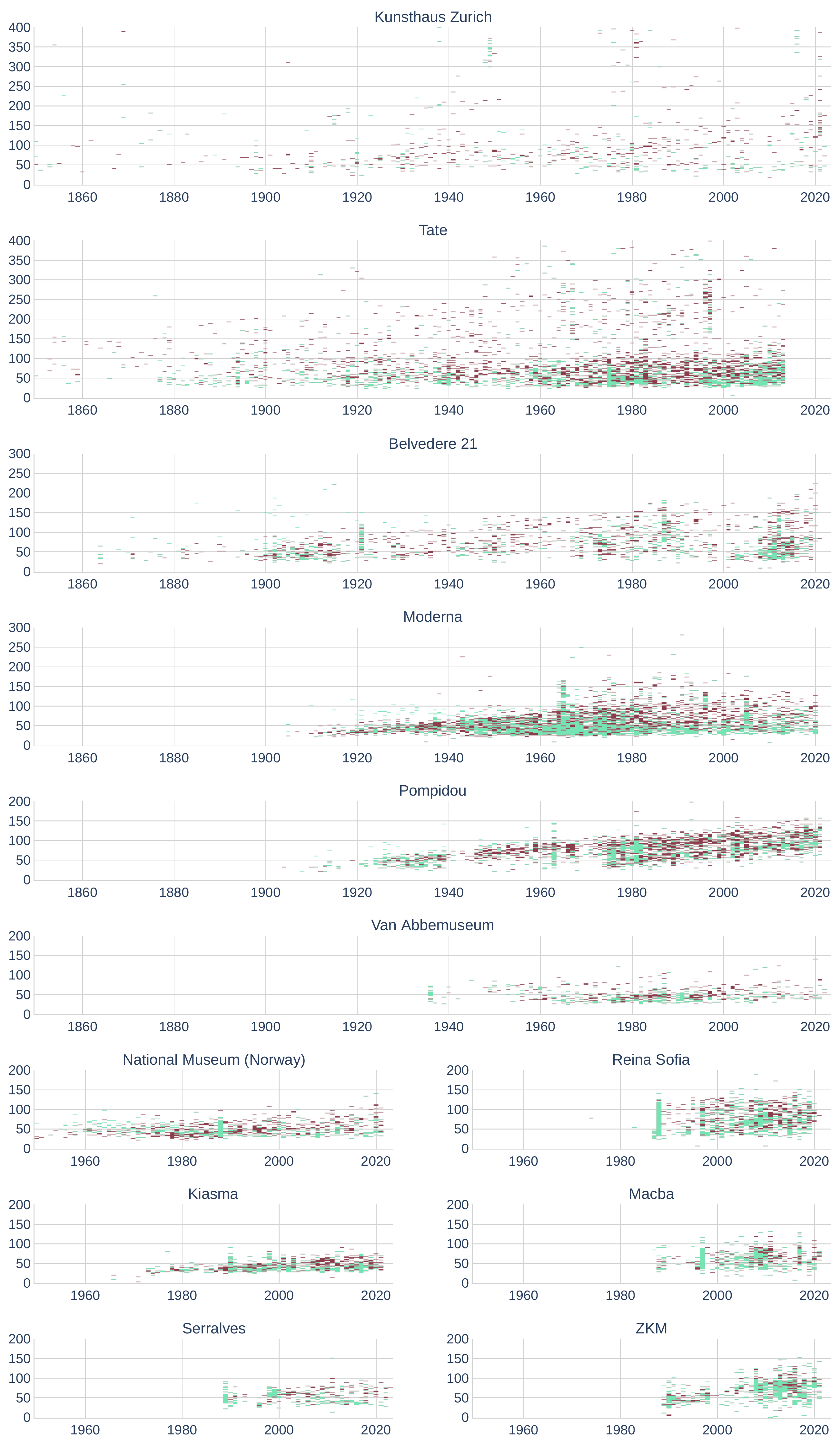}
 	\caption{Artist age for first (turquoise) and repeat (red) acquisitions  in 12 European museums over time.}
  \label{fig_ageAbsoluteTime}
\end{figure}

For further investigation of patterns and strategies of museum’s acquisition policies, we aggregated the data by year starting from the earliest acquisition record and by artist age. As a result, we obtained two-dimensional matrices, visualized as heatmaps in Figure \ref{fig_ageAbsoluteTime}. Turquoise color represents a number of first-time artist acquisitions in a particular museum, dark red color stands for recurring acquisitions, i.e. acquisitions of artists already in the collection. 

Plotting artist age over years for all artworks indicates cross-generational bursts and systematic trends in museum acquisitions. In case of Moderna - Sweden, we can see how a museum acquires artworks across the lifetime of artists, and the acquisition process is oriented towards the present, introducing artists in the collection mostly in the first 50 years of their life. For The Centre Pompidou and Tate repeating acquisitions are dominant for older artists and often brought to the collection after the death of the artist. We also note periods of exclusive repeating acquisitions for The Centre Pompidou in the 1950s and 1970s. 

Another systematic pattern in Figure \ref{fig_ageAbsoluteTime} is the introdution of several artists into the collection at once. This pattern expresses itself as turquoise vertical bursts (first acquisitions for various artists ages) and can be seen in the moment of forming a collection (Kiasma, Museo Reina Sofia, MACBA and Serralves) but also throughout time, for example with the collection expansion of Moderna - Sweden in the 1960s, or with recognition of media and media art as a part of contemporary art collection for ZKM in the 2010s. Yet another clear pattern, not found in all museums, is the clear upward trend in The Centre Pompidou acquisition policy, where the collection artist acquisition age seems to grow gradually in a systematic way. This paper we can only provide initial evidence for such patterns. Yet we can clearly see emerging quantitative patterns, including bursts, periodic episodes, and gradual trends, which are worthy of further investigation. 

\subsection{Artist age over artwork age }
\begin{figure*}
 	\noindent
 	\includegraphics[width=\textwidth]{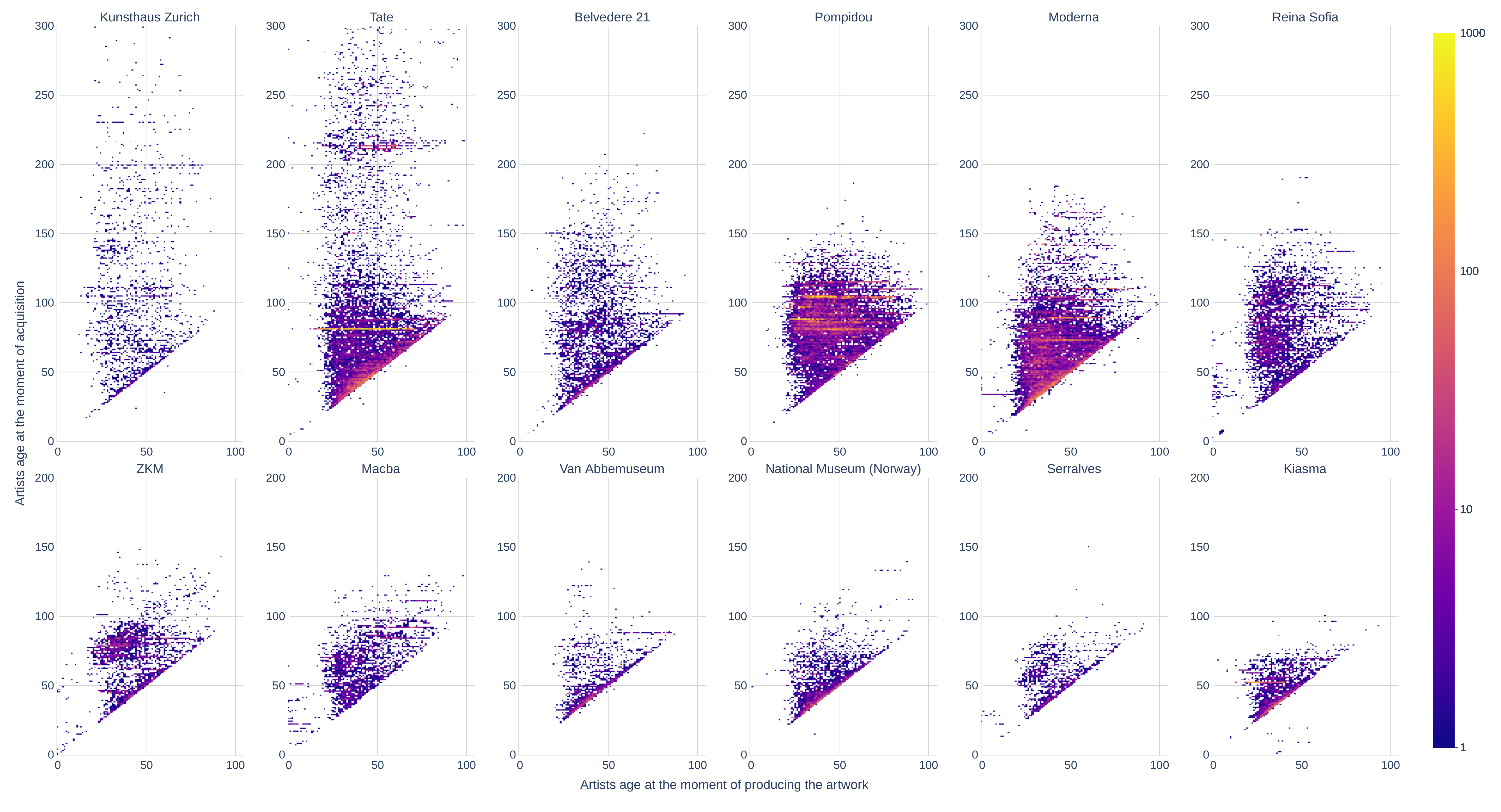}
 	\caption{Number of works (color) with artist age at acquisition over artist age at artwork creation in 12 European museums.  
 	}
  \label{fig_matrixMuseums}
\end{figure*}

To study support of living artists across museums we compared artist age at the moment of producing the artwork with the artist age at the moment of acquisition for the entire digitized collection for each museum. The results shown in Figure \ref{fig_matrixMuseums} reveal interesting structural differences across museums. Each cell in the heatmap matrices indicates the number of artworks acquired (y) at a given age of creation (x). There are several emerging patterns that are salient: artworks are mostly above the diagonal, i.e. preduced before acquisition, most museums acquire mid career close to production, i.e. close to the diagonal.  %

In general, there is a fade-off of acquisition probability with artist's age, only two out of 12 museums acquire artworks more than 200 years after their creation. And some of the others, including Kiasma, National Museum - Norway, and Serralves, have a clear focus on expanding their collections by acquiring from truly contemporary artists. Some museums, e.g. Tate and Moderna, tend to obtain artworks at the moment close to their creation (high values are alongside the diagonal, marked magenta to yellow). In some museums, there is a second wave of acquisition around the end of artist life-expectancy (diagonal density parallel above the diagonal). This second late-career or after-death acquisition pattern can be more (ZKM) or less dense (Serralves and MACBA).

In some museums, one can find strong horizontal signals apparently indicating whole oeuvre acquisitions for specific artists. This pattern can be seen for The Centre Pompidou and Moderna - Sweden as a burst in acquisitions for particular age (magenta lines), but it is also present in other museums as an increase in density (deep blue horizontal lines in Tate, Kunsthaus Zürich, and Van Abbemuseum). It is important to note that there is no clear vertical pattern present in the data. It means that museums do not show preference in acquiring artworks from any period of an artist's career. However, some museums like Belvedere 21 and Museo Reina Sofia demonstrate a tendency to include more works from earlier periods of the artist's lifetime (density area at the left side of plots). And other museums that expand their collection by acquiring artworks after the death of the artist (Kunsthaus Zürich and Tate) lean towards artworks produced before the artist reached 50. This can perhaps be explained by the life expectancy active career periods. In sum, through the construction of comparative heatmap matrices plotting acquisition age over creation age matrices, we have discovered systematically emerging structural patterns in acquisition habits or policies across museums. Srtikingly, despite the presence of some general patterns in acquisition strategies, as evident in the general fade-off to the past, the diagonal and horizontal structures, we also note that the patterns are verying across museums. Perhaps, seemingly trivial, we may state, that every museum adapts and modifies acquisition strategies to fulfill its own goal. More profound, we can take home that museum acquisition is a complex process in the sense of complex systems science \parencite{anderson1972more, IbrusSchichTamm2022}, where new previously hidden forms of quality emerges from the quantification of granular detail.

\section{Discussion}
This paper analyses a comprehensive set of European contemporary museum collections, employing a multi-scale approach to facilitate an in-depth exploration of emerging patterns across the diverse institutions under investigation. The research findings of this study reveal that each museum possesses a complex distinctive structure, rooted in heterogeneous yet resonant patterns of acquisition, as visually depicted in Figure \ref{fig_matrixMuseums}. Although these results may be anticipated by qualiative experts to some degree, the study unveils the intricate nature of the process over time, shaped by the interactions of numerous actors within the cultural landscape, institutions, and historical events. We believe that this study and a stream of follow-up research will contribute to a deeper understanding of the dynamics of art collection in artists, scholars, and museum stakeholders, including boards of trustees, directors, curators, donors, and visitors. Our visualization provide a holistic perspective on the phenomenon museum acquisition, offering a comprehensive overview of its complexities.

Some stunning results are that all collections show a remarkably close mean age in production and displaying a similarly centered unimodal distribution of ages. Moreover, if we observe an 8-year difference in the means of their production ages, with the lowest museum recording a mean age production of 34 and the highest museum showing a mean age of 41. It is evident that the discrepancy in mean acquisition age is significantly greater, amounting to 33 years. Moreover the study unveils that the mean acquisition age of artists is increasing across Europeans museums. 

Directors and curators may find our visualizations useful to assess and improve their museum's collection policies and acquisition methods. Building a collection, museums follow different priorities: emphasizing collection strengths, filling gaps, or acquiring works that will completely alter the institution’s collection profile \parencite{miller2022museum}. In a traditional sense, when a museums' mission is to preserve artifacts for future generations \parencite{jankauskas2022strategies, passebois2004building, jeffers2003museum}, collecting artworks from older or deceased artists might look as a way to ensure a higher quality of a collection. Yet given such an environment, an higher uncertainty level and lower earnings would force artists to get additional income sources, or look for other ways to sustain their work \parencite{menger1999artistic}. Due to the inherent specificity of artistic labor, artistic careers in such an environment would tend to be less financially stable and sparking promising artists to change career paths \parencite{alper2006artists}. Meanwhile, our work shows that not all museums acquire artworks after 25 years of artistic careers, as it stated by former Tate’s director Bowness \parencite{bowness_conditions_1990}. Many museums (Kiasma, National Museum - Norway, Serralves) acquire artworks many years earlier, thus actively participating in validating artistic careers. This phenomenon might be explained by the nature of contemporary painting and the market for contemporary art; painters in the 21st century might be recognized by community earlier \parencite{galenson2000careers}, countries where museums focus on conservation of older works should be aware that such cutting-edge collecting museums exist.

\section{Conclusion}

In this study we discovered diverse patterns in the museum acquisition of contemporary art. Some museums acquire works from renowned authors, while others support emerging talent. The mean museum collection lag between the creation of an artwork and its acquisition by European museums, as analyzed here, ranges from 3 to 33 years, which suggests that museums frequently recognize artists during their lifetime. Some museums adopt a more conservative strategy, reaping the fruits of long-term endeavors posthumously. The lag highlights how crucial it is for artists to remain dedicated to their work throughout their lives in order to receive recognition later in their careers or even after their passing.

\section*{Author contributions, acknowledgments and funding} 

 M.C., A.K., K.M. and M.S. contributed to the research design and co-wrote the manuscript. M.C., A.K and K.M. collected data. M.C. and A.K. contributed equally to this work as first authors. The authors thank Ederi Ojasoo, Tillmann Ohm, Mila Oiva, Mikhail Tamm, and the entire CUDAN team for useful conversations and input. All authors are supported by the CUDAN ERA Chair project, funded through the European Union's Horizon 2020 research and innovation program (Grant No. 810961)

\begingroup
\setlength{\emergencystretch}{8em}
\printbibliography
\FloatBarrier
\endgroup

\end{document}